\documentclass[aps,pra,preprintnumbers,showpacs,tightenlines]{revtex4}

\usepackage{amssymb}
\usepackage{amsmath}
\usepackage{graphicx}
\usepackage{epsfig}
\usepackage{subfigure}
\usepackage{amsfonts}
\usepackage{CJK}

\begin{document}

\title{Preparation of $n$-qubit Greenberger-Horne-Zeilinger entangled states in
cavity QED: An approach with tolerance to nonidentical qubit-cavity coupling constants}
\author{Chui-Ping Yang$^{1,2}$}

\address{$^1$Department of Physics, Hangzhou Normal University,
Hangzhou, Zhejiang 310036, China}

\address{$^2$State Key Laboratory of Precision Spectroscopy, Department of Physics,
East China Normal University, Shanghai 200062, China}

\date{\today}

\begin{abstract}
We propose a way for generating $n$-qubit Greenberger-Horne-Zeilinger (GHZ)
entangled states with a three-level qubit system and $\left( n-1\right) $
four-level qubit systems in a cavity. This proposal does not
require identical qubit-cavity
coupling constants, and thus is tolerant to qubit-system parameter
nonuniformity and nonexact placement of qubits in a cavity. The proposal
does not require adjustment of the qubit-system level
spacings during the entire operation. Moreover, it is shown that
entanglement can be deterministically generated using this method and the
operation time is independent of the number of qubits. The present proposal is
quite general, which can be applied to physical systems such as various
types of superconducting devices coupled to a resonator or atoms trapped in
a cavity.
\end{abstract}

 \maketitle
\date{\today}

\begin{center}
\textbf{I. INTRODUCTION}
\end{center}

Quantum entanglement plays a crucial role in quantum information processing.
Entanglement of ten photons [1], eight ions [2], three spins [3], two atoms
in microwave cavity QED [4], or two excitons in a single quantum dot [5] has
been demonstrated in experiments. In addition, experimental preparation of
three-qubit entanglement with superconducting qubits or a superconducting
qubit coupled to two microscopic two-level systems has been reported
recently [6-8]. Although multi-particle entanglement was experimentally
created in photons and trapped ions, it is still greatly challenging to
create multi-qubit entanglement in other important physical systems.

As is well known, multi-qubit GHZ (Greenberger-Horne-Zeilinger) entangled
states are of great interest in the foundations of quantum mechanics and
measurement theory, and significant in quantum information processing [9],
quantum communication [10-12], error correction protocols [13], and
high-precision spectroscopy [14]. Over the past ten years, based on cavity
QED technique, many different theoretical methods for creating multi-qubit
GHZ entangled states with atoms and superconducting qubits have been
presented [15-25]. For instances, (i) the proposals in [15-17] for
implementing a GHZ entangled state are based on identical qubit-cavity
coupling constants; (ii) the approaches in [18-20] are aimed at
probabilistic generation of a GHZ state; (iii) the method presented in [17]
is based on the use of an auxiliary qubit, measurement on the qubit states,
and adjustment of the qubit level spacings during the entire operation; (iv)
the proposals in [21,22] are based on photon detection outside the cavity;
and (v) the proposals in [23-25] are based on sending qubits (i.e., atoms)
through a cavity. These proposals are important because they opened new
avenues for creating multi-particle entanglement.

In this paper, we will focus on a situation that the qubit-cavity coupling
constants are nonidentical. This situation often exists in superconducting
qubits coupled to a cavity or a resonator. For solid-state devices, the
device parameter nonuniformity is often a problem, which results in
nonidentical qubit-cavity coupling constants though qubits are placed at
locations of a cavity where the magnetic fields or the electric fields of
the cavity mode are the same. In addition, the nonidentical qubit-cavity
coupling constants may also result from nonexact placement of qubits in a
cavity. In the following, our goal is wish to present a way for
deterministic preparation of a $n$-qubit GHZ state with tolerance to the
qubit-system parameter nonuniformity and nonexact placement of qubits in a
cavity. As shown below, this proposal also has these advantages: (i) The
operation time is independent of the number of qubits in the cavity; (ii)
When compared with the approach in [17], no auxiliary qubits, no measurement
on the qubit states, and no adjustment of the qubit level spacings during
the entire operation is needed (note that adjustment of the level spacings
of the qubits during the operation is not desired in experiments and may
cause extra errors); (iii) No adjustment of the cavity mode frequency is
required during the entire operation; and (iv) There is no need of photon
detection. This proposal is quite general, which can be applied to various
types of superconducting qubits and atoms trapped in a cavity.

This paper is organized as follows. In Sec.~II, we briefly review the basic
theory of a three-level quantum system or four-level quantum systems coupled
to a single-mode cavity and/or driven by classical pulses. In Sec.~III, we
show how to generate an $n$-qubit GHZ state with one three-level quantum
system and $\left( n-1\right) $ four-level quantum systems in a cavity. In
Sec.~IV, we compare our proposal with the previous ones. In Sec. V, we give
a brief discussion of the experimental issues and possible experimental
implementation with superconducting qubits coupled to a resonator. A
concluding summary is given in Sec. VI.

\begin{center}
\textbf{II. BASIC THEORY}
\end{center}

In this section, we will introduce three types of interaction of qubit
systems with the cavity mode and/or the pulse. The results presented below
will be employed for generation of a multi-qubit GHZ state discussed in next
section.
\begin{figure}[tbp]
\includegraphics[bb=76 94 424 682, width=8.6 cm, clip]{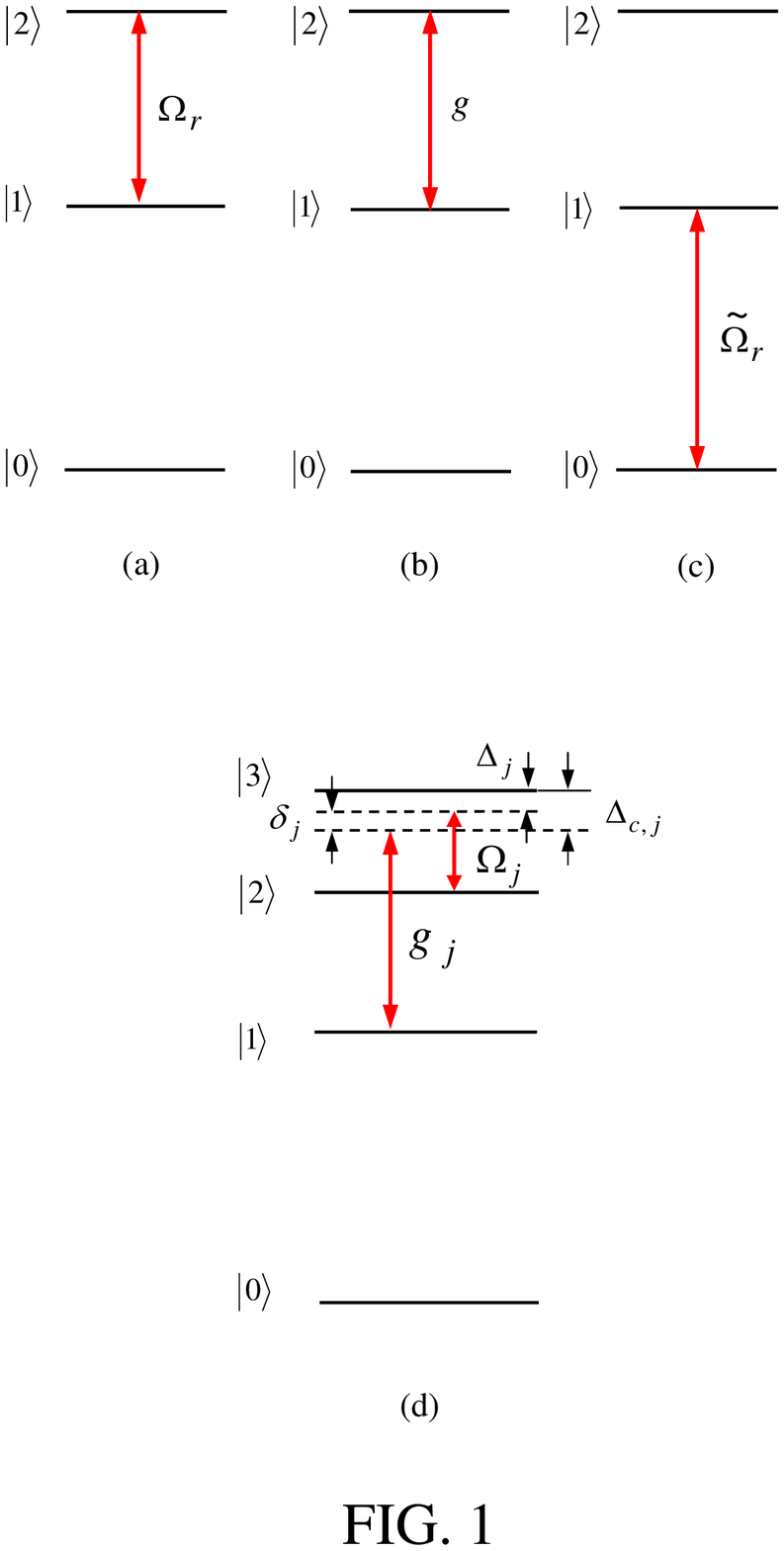} %
\vspace*{-0.08in}
\caption{(Color online) (a) and (c) System-pulse resonant interaction for
qubit system 1. In (a), the pulse is resonant with the $\left|1\right\rangle
\leftrightarrow \left| 2\right\rangle $ transition; while in (c) the pulse
is resonant with the $\left| 0\right\rangle \leftrightarrow \left|
1\right\rangle $ transition. (b) System-cavity resonant interaction for
qubit system 1. The pulse is resonant with the $\left| 1\right\rangle
\leftrightarrow \left| 2\right\rangle $ transition. (d) System-cavity-pulse
off-resonant Raman coupling for $n-1$ qubit systems ($2,3,...,n$) in a
cavity. For simplicity, we here only draw a figure for qubit system $j$
interacting with the cavity mode and a classical pulse ($j=2,3,...,n$). $%
\delta _j=\Delta _{c,j}-\Delta _j$ is the detuning of the cavity mode with
the pulse, $\Delta _j=\omega _{31}^j-\omega_j $ is the detuning between the
pulse frequency $\omega_j $ and the $\left| 1\right\rangle \leftrightarrow
\left| 3\right\rangle $ transition frequency $\omega _{31}^j$ of qubit
system $j.$ Note that $g_j$, $\delta _j,$ $\Delta _{c,j},$ and $\Delta _j$
may be different for qubit systems ($2,3,...,n$) due to nonidentical level
spacings of the qubit systems which are caused by the nonuniformity of the
qubit system parameters. The coupling constant $g_j$ may also vary with
qubits due to nonexact placement of qubits in a cavity. The Rabi frequency
of the pulse applied to qubit system $j$ is denoted by $\Omega _j$. }
\label{fig:1}
\end{figure}

\begin{center}
\textbf{A. System-pulse resonant interaction}
\end{center}

Consider a three-level qubit system (say , qubit system 1) driven by a
classical pulse. Suppose that the pulse is resonant with the transition $%
\left| 1\right\rangle \leftrightarrow \left| 2\right\rangle $ of the qubit
system 1 but decoupled from the transition between any two other levels
[Fig.~1(a)]. The interaction Hamiltonian in the interaction picture is given
by
\begin{equation}
H_I=\hbar \left( \Omega _re^{i\phi }\left| 1\right\rangle \left\langle
2\right| +\text{H.c.}\right) ,
\end{equation}
where $\Omega _r$ and $\phi $ are the Rabi frequency and the initial phase
of the pulse, respectively. Based on the Hamiltonian (1), it is
straightforward to show that a pulse of duration $t$ results in the
following rotation
\begin{eqnarray}
\left| 1\right\rangle &\rightarrow &\cos \Omega _rt\left| 1\right\rangle
-ie^{-i\phi }\sin \Omega _rt\left| 2\right\rangle ,  \nonumber \\
\left| 2\right\rangle &\rightarrow &-ie^{i\phi }\sin \Omega _rt\left|
1\right\rangle +\cos \Omega _rt\left| 2\right\rangle .
\end{eqnarray}
Note that the resonant interaction can be done within a very short time by
increasing the pulse Rabi frequency $\Omega _r$ (i.e., via increasing the
pulse intensity).

In the following, we also need the resonant interaction of the pulse with
the $\left| 0\right\rangle \leftrightarrow \left| 1\right\rangle $ transiton
of the qubit system 1 [Fig. 1(c)]. In this case, we have
\begin{eqnarray}
\left| 0\right\rangle &\rightarrow &\cos \widetilde{\Omega }_rt\left|
0\right\rangle -ie^{-i\phi }\sin \widetilde{\Omega }_rt\left| 1\right\rangle
,  \nonumber \\
\left| 1\right\rangle &\rightarrow &-ie^{i\phi }\sin \Omega _rt\left|
0\right\rangle +\cos \Omega _rt\left| 1\right\rangle ,
\end{eqnarray}
where $\widetilde{\Omega }_r$ is the Rabi frequency of the pulse.

\begin{center}
\textbf{B. System-cavity resonant interaction}
\end{center}

Consider qubit system 1 coupled to a single-mode cavity field. Suppose that
the cavity mode is resonant with the $\left| 1\right\rangle \leftrightarrow
\left| 2\right\rangle $ transition while decoupled (highly detuned) from the
$\left| 0\right\rangle \leftrightarrow \left| 2\right\rangle $ transition
and the $\left| 0\right\rangle \leftrightarrow \left| 1\right\rangle $
transition of the qubit system 1[Fig.~1(b)]. The interaction Hamiltonian in
the interaction picture, after the rotating-wave approximation, is described
by
\begin{equation}
H_I=\hbar \left( ga^{+}\left| 1\right\rangle \left\langle 2\right| +\text{%
h.c.}\right) ,
\end{equation}
where $a^{+}$ and $a$ are the creation and annihilation operators of the
cavity mode, and $g$ is the coupling constant between the cavity mode and
the $\left| 1\right\rangle \leftrightarrow \left| 2\right\rangle $
transition of the qubit system. Under the Hamiltonian (4), the time
evolution of the states $\left| 2\right\rangle \left| 0\right\rangle _c$ and
$\left| 1\right\rangle \left| 1\right\rangle _c$ of the whole system are as
follows
\begin{eqnarray}
\left| 2\right\rangle \left| 0\right\rangle _c &\rightarrow &\cos gt\left|
2\right\rangle \left| 0\right\rangle _c-i\sin gt\left| 1\right\rangle \left|
1\right\rangle _c,  \nonumber \\
\left| 1\right\rangle \left| 1\right\rangle _c &\rightarrow &-i\sin gt\left|
2\right\rangle \left| 0\right\rangle _c+\cos gt\left| 1\right\rangle \left|
1\right\rangle _c,
\end{eqnarray}
where the states $\left| 0\right\rangle _c$ and $\left| 1\right\rangle _c$
are the cavity-mode vacuum state and single-photon state, respectively.

\begin{center}
\textbf{C. System-cavity-pulse off-resonant Raman coupling}
\end{center}

Consider $\left( n-1\right) $ four-level qubit systems ($2,3,...,n$). The
cavity mode is coupled to the $\left| 1\right\rangle \leftrightarrow \left|
3\right\rangle $ transition of each qubit system, but decoupled (highly
detuned) from the transition between any other two levels [Fig.~1(d)]. In
addition, a classical pulse is applied to each one of qubit systems ($%
2,3,...,n$), which is coupled to the $\left| 2\right\rangle \leftrightarrow
\left| 3\right\rangle $ transition but decoupled from the transition between
any other two levels [Fig.~1(d)]. In the interaction picture, the
Hamiltonian for the whole system is
\begin{equation}
H=\sum_{j=2}^n\left[ \hbar g_j(e^{-i\Delta _{c,j}t}a^{+}\sigma _{13,j}^{-}+%
\text{H.c.})+\hbar \Omega _j(e^{-i\Delta _jt}\sigma _{23.j}^{-}+\text{H.c.}%
)\right] ,
\end{equation}
where the subscript $j$ represents the $j$th qubit system$,$ $\Omega _j$ is
the Rabi frequency of the pulse applied to the $j$th qubit system, $\sigma
_{13,j}^{-}=\left| 1\right\rangle _j\left\langle 3\right| ,$ $\sigma
_{23,j}^{-}=\left| 2\right\rangle _j\left\langle 3\right| ,$ and $g_j$ is
the coupling constant between the cavity mode and the $\left| 1\right\rangle
\leftrightarrow \left| 3\right\rangle $ transition of the $j$th qubit system.

The detuning between the $\left| 2\right\rangle \leftrightarrow \left|
3\right\rangle $ transition frequency $\omega _{32}^j$ of the $j$th qubit
system and the frequency $\omega _j$ of the pulse applied to the $j$th qubit
system is $\Delta _j=\omega _{32}^j-\omega _j$ ($j=2,3,...,n$) [Fig. 1(d)].
In addition, the detuning between the $\left| 1\right\rangle \leftrightarrow
\left| 3\right\rangle $ transition frequency $\omega _{31}^j$ of the $j$th
qubit system and the cavity-mode frequency $\omega _c$ is $\Delta
_{c,j}=\omega _{31}^j-\omega _c$ [Fig.~1(d)]. Note that the detuning $\Delta
_{c,j}$ is not the same for each qubit system (i.e., dependent of $j$) in
the case when the level spacings are nonidentical for each qubit system.
Under the condition $\Delta _{c,j}\gg g_j$ and $\Delta _j\gg \Omega _j,$ the
level $\left| 3\right\rangle $ can be adiabatically eliminated [26] and the
effective Hamiltonian is thus given by [27-29]

\begin{eqnarray}
H_{\mathrm{eff}} &=&-\hbar \sum_{j=2}^{n+1}\left[ \frac{\Omega _j^2}{\Delta
_j}\left| 2\right\rangle _j\left\langle 2\right| +\frac{g_j^2}{\Delta _{c,j}}%
a^{+}a\left| 1\right\rangle _j\left\langle 1\right| \right.  \nonumber \\
&&\ \ \ \ \left. +\chi _j(e^{-i\delta _jt}a^{+}\sigma _{12,j}^{-}+\text{H.c.}%
)\right] ,
\end{eqnarray}
where $\sigma _{12,j}^{-}=\left| 1\right\rangle _j\left\langle 2\right| ,$ $%
\chi _j=\frac{\Omega _jg_j}2(1/\Delta _j+1/\Delta _{c,j})$, and
\begin{equation}
\delta _j=\Delta _{c,j}-\Delta _j=\omega _{31}^j-\omega _{32}^j-\omega
_c+\omega _j.
\end{equation}
From Eq.~(8), it can be seen that the detuning $\delta _j$ is adjustable by
changing the pulse frequency $\omega _j.$ With suitable choice of the pulse
frequencies, we can satisfy
\begin{equation}
\delta _2=\delta _3=...=\delta _n=\delta ,
\end{equation}
which will apply below.

For $\delta \gg g_j^2/\Delta _{c,j},$ $\Omega _j^2/\Delta _j,$ $\chi _j,$
there is no energy exchange between the qubit systems and the cavity mode.
Thus, under the condition (9), the effective Hamiltonian (7) can be written
as [25,30,31]
\begin{eqnarray}
H_{\mathrm{eff}} &=&-\hbar \sum_{j=2}^n\left( \frac{\Omega _j^2}{\Delta _j}%
\left| 2\right\rangle _j\left\langle 2\right| +\frac{g_j^2}{\Delta _{c,j}}%
a^{+}a\left| 1\right\rangle _j\left\langle 1\right| \right)  \nonumber \\
&&\ \ \ \ \ -\hbar \sum_{j=2}^n\left[ \frac{\chi _j^2}\delta \left(
a^{+}a\left| 1\right\rangle _j\left\langle 1\right| -aa^{+}\left|
2\right\rangle _j\left\langle 2\right| \right) \right]  \nonumber \\
&&\ \ \ \ \ +\hbar \sum_{j\neq j^{\prime }=2}^n\frac{\chi _j\chi _{j^{\prime
}}}\delta \left( \sigma _{12,j}^{+}\sigma _{12,j^{\prime }}^{-}+\sigma
_{12,j}^{-}\sigma _{12,j^{\prime }}^{+}\right) ,  \nonumber \\
&&
\end{eqnarray}
where the two terms in the second line above describe the photon-number
dependent Stark shifts induced by the off-resonant Raman coupling, and the
two terms in the last parentheses describe the ``dipole'' coupling between
the two qubit systems ($j,j^{\prime })$ mediated by the cavity mode and the
classical pulses. In the case when the level $\left| 2\right\rangle $ of
each qubit system is not populated, the Hamiltonian (10) reduces to
\begin{equation}
H_{\mathrm{eff}}=-\hbar \sum_{j=2}^n\left( \frac{g_j^2}{\Delta _{c,j}}+\frac{%
\chi _j^2}\delta \right) a^{+}a\left| 1\right\rangle _j\left\langle 1\right|
.
\end{equation}
It is easy to see that the states $\left| 0\right\rangle _j\left|
0\right\rangle _c,$ $\left| 1\right\rangle _j\left| 0\right\rangle _c,$ and $%
\left| 0\right\rangle _j\left| 1\right\rangle _c$ remain unchanged under the
Hamiltonian (11). However, if the cavity mode is initially in the photon
state $\left| 1\right\rangle _c,$ the time evolution of the state $\left|
1\right\rangle _j$ of the $j$th qubit system under the Hamiltonian (11) is
given by
\begin{equation}
\left| 1\right\rangle _j\left| 1\right\rangle _c\rightarrow e^{i\lambda
_jt}\left| 1\right\rangle _j\left| 1\right\rangle _c,
\end{equation}
where $\lambda _j=\frac{g_j^2}{\Delta _{c,j}}+\frac{\chi _j^2}\delta ,$
which can be further written as
\begin{equation}
\lambda _j=\frac{g_j^2}{\Delta _{c,j}}+\frac{\Omega _j^2g_j^2}{4\delta }%
\left( \frac 1{\Delta _j}+\frac 1{\Delta _{c,j}}\right) ^2.
\end{equation}
Note that the parameters $g_j$ is not adjustable once the qubit systems
(e.g., solid-state devices) are designed and built in a cavity, and the
detuning $\Delta _{c,j}=\omega _{31}^j-\omega _c$ is fixed when the cavity
mode frequency is chosen. As can be seen in Eq.~(13), the parameter $\lambda
_j$ here is adjustable by changing the pulse Rabi frequency $\Omega _j.$

\begin{center}
\textbf{III. GENERATION OF MULTI-QUBIT GHZ STATES}
\end{center}

Let us consider $n$ qubit systems ($1,2,...,n$) in a single-mode cavity. The
qubit system $1$ has three levels shown in Fig.~1(a,b,c) while the four
levels of the qubit systems ($2,3,...,n$) are depicted in Fig.~1(d). For
qubit system $1,$ the cavity mode is resonant with the $\left|
1\right\rangle \leftrightarrow \left| 2\right\rangle $ transition but highly
detuned from the transition between any two other levels [Fig.~1(b)]. In
contrast, for qubit systems ($2,3,...,n$), the cavity mode is off-resonant
with the $\left| 1\right\rangle \leftrightarrow \left| 3\right\rangle $
transition but highly detuned from the transition between any other two
levels [Fig. 1(d)]. These requirements can be achieved by an appropriate
choice of qubit systems (e.g., atoms), prior adjustment of the level
spacings of the qubit systems (e.g., superconducting devices), or prior
adjustment of the cavity mode frequency before the operation. Note that the
cavity mode frequency for both optical cavities and microwave cavities can
be changed in various experiments (e.g., see, [32-36]). And, for
superconducting qubit systems, the level spacings can be readily adjusted by
varying the external parameters (e.g., the external magnetic flux and gate
voltage for superconducting charge-qubit systems, the current bias or flux
bias in the case of superconducting phase-qubit systems and flux-qubit
systems, see e.g. [37-39]).

Suppose that the cavity mode and each of qubit systems ($1,2,...,n$) are
initially in $\left| 0\right\rangle _c$ and $\left( \left| 0\right\rangle
+\left| 1\right\rangle \right) /\sqrt{2},$ respectively. The initial state
for each of the qubit systems here can be easily prepared by the application
of classical pulses. The whole procedure for preparing qubit systems ($%
1,2,...,n$) in a GHZ state is shown as follows:

Step (i): Apply a classical pulse (with a frequency $\omega =\omega _{21}$
and $\phi =-\pi /2$) to qubit system $1$ for a duration $t_{1a}=\pi /\left(
2\Omega _r\right) $ [Fig. 1(a)]$,$ wait to have the cavity mode resonantly
interacting with the $\left| 1\right\rangle \leftrightarrow \left|
2\right\rangle $ transition of the qubit system $1$ for a time interval $%
t_{1b}=\pi /(2g)$ [Fig. 1(b)]$,$ and then apply a pulse (with a frequency $%
\omega =\omega _{10}$ and $\phi =-\pi /2$) to qubit system $1$ for a
duration $t_{1c}=\pi /(2\widetilde{\Omega }_r)$ [Fig. 1(c)]. According to
Eqs. (2), (3), and (5), it can be seen that after the operation of this
step, the following transformation is obtained:
\begin{equation}
\begin{array}{c}
\left| 0\right\rangle _1\left| 0\right\rangle _c \\
\left| 1\right\rangle _1\left| 0\right\rangle _c
\end{array}
\stackrel{t_{1a}}{\longrightarrow }
\begin{array}{c}
\left| 0\right\rangle _1\left| 0\right\rangle _c \\
\left| 2\right\rangle _1\left| 0\right\rangle _c
\end{array}
\stackrel{t_{1b}}{\longrightarrow }
\begin{array}{c}
\left| 0\right\rangle _1\left| 0\right\rangle _c \\
-i\left| 1\right\rangle _1\left| 1\right\rangle _c
\end{array}
\stackrel{t_{1c}}{\longrightarrow }
\begin{array}{c}
\left| 1\right\rangle _1\left| 0\right\rangle _c \\
i\left| 0\right\rangle _1\left| 1\right\rangle _c
\end{array}
,
\end{equation}
which leads the initial state $\prod_{j=1}^n\left( \left| 0\right\rangle
_j+\left| 1\right\rangle _j\right) \left| 0\right\rangle _c$ of the whole
system to the following state:
\begin{equation}
\prod_{j=2}^n\left( \left| 0\right\rangle _j+\left| 1\right\rangle _j\right)
\left( \left| 1\right\rangle _1\left| 0\right\rangle _c+i\left|
0\right\rangle _1\left| 1\right\rangle _c\right) .
\end{equation}
Here and below, the normalization factor $1/2^{n/2}$ is omitted for
simplicity. Eq.~(15) shows that the levels $\left| 1\right\rangle $ and $%
\left| 2\right\rangle $ of qubit system $1$ are not populated in the case
when the cavity mode is in the single-photon state $\left| 1\right\rangle _c$%
. Hence, after the operation of this step, the qubit system $1$ is decoupled
from the cavity mode during the operation of next step.

Step (ii): Apply a classical pulse (with a duration $t_2$) to each of qubit
systems ($2,3,...,n$) to induce the off-resonant Raman coupling described in
Sec.~II C [Fig.~1(d)]. By adjusting the pulse frequencies, we set $\delta
_2=\delta _3=...=\delta _n=\delta .$ Since the level $\left| 2\right\rangle $
for each of qubit systems ($2,3,...,n$) is not populated during the
operation of step (i) above, the effective Hamiltonian describing this step
of operation is given by Eq.~(11). Accordingly, when the cavity mode is in
the single-photon state $\left| 1\right\rangle _c,$ the time evolution of
the state $\left| 1\right\rangle _j$ for qubit system $j$ is then given by
Eq.~(12). Set $\lambda _2=\lambda _3=\cdot \cdot \cdot =\lambda _n=\lambda ,$
which can be readily achieved by adjusting the Rabi frequencies $\Omega _2,$
$\Omega _3,...,\Omega _n$ of the pulses applied to qubit systems ($2,3,...,n$%
). From Eq.~(12), one can see that for $t_2=\pi /\lambda $, we have the
transformation $\left| 1\right\rangle _j\left| 1\right\rangle _c\rightarrow
-\left| 1\right\rangle _j\left| 1\right\rangle _c$ ($j=2,3,...n$). Thus, the
state (15) becomes
\begin{equation}
\prod_{j=2}^n\left( \left| 0\right\rangle _j+\left| 1\right\rangle _j\right)
\left| 1\right\rangle _1\left| 0\right\rangle _c+i\prod_{j=2}^n\left( \left|
0\right\rangle _j-\left| 1\right\rangle _j\right) \left| 0\right\rangle
_1\left| 1\right\rangle _c,
\end{equation}
which shows that the $n$ qubit systems ($1,2,...n$) have been entangled to
each other after the above operations. Since the qubit systems ($1,2,...,n$)
are also entangled with the cavity mode, we will need to perform the
following operation to disentangle qubit systems ($1,2,...n$) from the
cavity mode.

Step (iii): Perform a reverse operation described in step (i). Namely, apply
a classical pulse (with a frequency $\omega =\omega _{10}$ and an initial
phase $\phi =\pi /2$) to qubit system $1$ for a duration $t_{1c}=\pi /\left(
2\widetilde{\Omega }_r\right) $ [Fig.~1(c)]$,$ wait to have the cavity mode
resonantly interacting with the $\left| 1\right\rangle \leftrightarrow
\left| 2\right\rangle $ transition of the qubit system $1$ for a time
interval $t_{1b}=\pi /(2g)$ [Fig. 1(b)]$,$ and then apply a pulse (with a
frequency $\omega =\omega _{21}$ and $\phi =\pi /2$) to qubit system $1$ for
a duration $t_{1a}=\pi /(2\Omega _r)$ [Fig. 1(a)]. According to Eqs. (2),
(3), and (5), it can be seen that after the operation of this step, we have:
\begin{equation}
\begin{array}{c}
\left| 1\right\rangle _1\left| 0\right\rangle _c \\
\left| 0\right\rangle _1\left| 1\right\rangle _c
\end{array}
\stackrel{t_{1c}}{\longrightarrow }
\begin{array}{c}
\left| 0\right\rangle _1\left| 0\right\rangle _c \\
-\left| 1\right\rangle _1\left| 1\right\rangle _c
\end{array}
\stackrel{t_{1b}}{\longrightarrow }
\begin{array}{c}
\left| 0\right\rangle _1\left| 0\right\rangle _c \\
i\left| 2\right\rangle _1\left| 0\right\rangle _c
\end{array}
\stackrel{t_{1a}}{\longrightarrow }
\begin{array}{c}
\left| 0\right\rangle _1\left| 0\right\rangle _c \\
i\left| 1\right\rangle _1\left| 0\right\rangle _c
\end{array}
,
\end{equation}
which leads the state (16) to the following state
\begin{equation}
\left[ \prod_{j=2}^n\left( \left| 0\right\rangle _j+\left| 1\right\rangle
_j\right) \left| 0\right\rangle _1-\prod_{j=2}^n\left( \left| 0\right\rangle
_j-\left| 1\right\rangle _j\right) \left| 1\right\rangle _1\right] \otimes
\left| 0\right\rangle _c.
\end{equation}
Eq.~(18) demonstrates that the cavity mode returns to its original vacuum
state and the qubit systems ($1,2,...,n$) have been disentangled from the
cavity mode after the above operations.

The left part of the product in the state (18) can be rewritten as
\begin{equation}
\ \ \left| 0\right\rangle _1\left| +\right\rangle _2\cdot \cdot \cdot \left|
+\right\rangle _n-\left| 1\right\rangle _1\left| -\right\rangle _2\cdot
\cdot \cdot \left| -\right\rangle _n,
\end{equation}
where $\left| +\right\rangle _j=\left| 0\right\rangle _j+\left|
1\right\rangle _j$, and $\left| -\right\rangle _j=\left| 0\right\rangle
_j-\left| 1\right\rangle _j$ ($j=2,3,...,n$). Since $\left| +\right\rangle
_j $ is orthogonal to $\left| -\right\rangle _j,$ the state (19) is a GHZ
entangled state of $n$ qubits.

To reduce the operation errors, the level-spacing inhomogeneity in each
four-level system needs to be larger than the bandwidth of the applied
pulse, such that the overlapping of pulse spectra or the transition between
any two irrelevant levels is negligible. This requirement can be achieved by
prior adjustment of the qubit level spacings. For superconducting qubit
systems, the level spacings can be readily adjusted by varying the external
parameters [37-39]. To simplify our presentation, we will not give a
detailed discussion here. Note that how to have the irrelevant levels not
affected by the pulses or the cavity mode via prior adjustment of the level
spacings was previously discussed (e.g., see [27,41]).

Several additional points need to be made, which are as follows:

(i) Since the cavity mode is off-resonant with the $\left| 1\right\rangle
\leftrightarrow \left| 3\right\rangle $ transition of qubit system $j$ ($%
j=2,3,...n$) during steps (i) and (iii), a phase shift $\exp (i\varphi _j)$
happens to the state $\left| 1\right\rangle $ of qubit system $j$ when the
cavity mode is in the single photon state $\left| 1\right\rangle _c$ for
either of steps (i) and (iii)$.$ Here, $\varphi _j=g_j^2\left(
t_{1b}+t_{1c}\right) /\Delta _{c,j}.$ It is easy to find that if this
unwanted phase shift for qubit system $j$ is considered, the fidelity of the
prepared GHZ state is given by
\begin{equation}
\emph{F}\simeq \frac 14\left[ 1+\prod_{j=2}^n\frac{\left( 1+e^{-i2\varphi
_j}\right) }2\right] \left[ 1+\prod_{j=2}^n\frac{\left( 1+e^{i2\varphi
_j}\right) }2\right] ,
\end{equation}
which shows that for $\varphi _j\sim 0,$ i.e., when the condition
\begin{equation}
g_j^2\left( t_{1b}+t_{1c}\right) /\Delta _{c,j}\ll 1
\end{equation}
is met$,$ we have $\emph{F}\sim 1.$ The condition (21) can be reached by
increasing $g$ and $\widetilde{\Omega }_r$ to shorten the operation time $%
t_{1b}+t_{1c}$ or increasing the ratio $\Delta _{c,j}/g_j^2.$

(ii) For steps (i) and (iii), the pulse process depicted in Fig.~1(a) and
(c) must be much faster than the process of the cavity mode resonantly
interacting with the $\left| 1\right\rangle \leftrightarrow \left|
2\right\rangle $ transition of the qubit system $1$ [Fig.~1(b)]$,$ i.e.,
\begin{equation}
t_{1a},t_{1c}\ll t_{1b,}
\end{equation}
such that the internal transition between the two levels $\left|
1\right\rangle $ and $\left| 2\right\rangle $ of qubit system $1$ during the
pulses, induced by the resonant interaction of the cavity mode with the $%
\left| 1\right\rangle \leftrightarrow \left| 2\right\rangle $ transition of
qubit system $1,$ is negligible. Note that the condition (22) can be readily
achieved by increasing the pulse Rabi frequencies $\Omega _r$ and $%
\widetilde{\Omega }_r$ such that $\Omega _r,$ $\widetilde{\Omega }_r\gg g.$

(iii) The level $\left| 2\right\rangle $ of qubit system $1$ is populated
for a time interval $t_{1a}+t_{1b}$ during step (i) or step (iii), thus the
condition
\begin{equation}
t_{1a}+t_{1b}\ll \gamma _{2r}^{-1},\gamma _{2p}^{-1}
\end{equation}
needs to be satisfied in order to reduce decoherence caused due to
spontaneous emission and dephasing of the level $\left| 2\right\rangle $ of
qubit system $1.$ Here, $\gamma _{2r}^{-1}$ and $\gamma _{2p}^{-1}$ are the
energy relaxation time and dephasing time of the level $\left|
2\right\rangle $ of qubit system $1,$ respectively.

From the above description, one can see that this proposal has the following
advantages:

(i) No identical qubit-cavity coupling constants for the $(n-1)$ qubit
systems ($2,3,...,n$) are required, thus, this proposal is tolerant to the
qubit-system parameter nonuniformity and nonexact placement of qubits in a
cavity;

(ii) No adjustment of the level spacings of the qubit systems or adjustment
of the cavity mode frequency during the entire operation is needed;

(iii) Neither auxiliary qubit systems nor measurement on the qubit states is
needed;

(iv) No photon detection is needed;

(v) The entanglement preparation is deterministic;

(vi) The entire operation time is given by
\begin{equation}
\tau =\pi /g+\pi /\Omega _r+\pi /\widetilde{\Omega }_r+\pi /\lambda ,
\end{equation}
which is independent of the number of qubits in the cavity.

In addition, for the description given above, it can be seen that:

(vi) During the entire operation, the levels $\left| 2\right\rangle $ and $%
\left| 3\right\rangle $ of qubit systems ($2,3,...,n$) are unpopulated and
thus decoherence due to spontaneous emission and dephasing from these levels
are greatly reduced;

(vii) The level $\left| 2\right\rangle $ of qubit system $1$ is only
populated for a very short time $t_{1a}+t_{1b}$ during step (i) or step
(iii); and

(viii) The level $\left| 2\right\rangle $ of qubit system $1$ is not
occupied during step (ii) [note that the operation of this step (ii)
requires much longer time $t_2$ than both step (i) and step (iii)].

Above we have discussed how to prepare a multi-qubit GHZ state with a
three-level qubit system and $\left( n-1\right) $ four-level qubit systems
in a cavity. The discussion given above is based on qubit systems for which
the level spacings become narrower as the levels go up [see Fig. 1(a,b,c)
for qubit system 1 and Fig.~1(d) for qubit systems ($2,3,...,j$)]. Note that
this limitation is unnecessary. Namely, this proposal is also applicable to:
(i) the qubit system $1$ with such three levels, for which the level spacing
between the two lowest levels $\left| 0\right\rangle $ and $\left|
1\right\rangle $ is smaller than that between the two upper levels $\left|
1\right\rangle $ and $\left| 2\right\rangle ;$ and (ii)\ the $\left(
n-1\right) $ qubit systems ($2,3,...,n$) with these four levels, for which
the level spacing between the two levels $\left| i\right\rangle $ and $%
\left| i+1\right\rangle $ is larger or smaller than that between the two
levels $\left| i+1\right\rangle $ and $\left| i+2\right\rangle $ (here, $%
i=0,1$). Furthermore, since the level $\left| 0\right\rangle $ of each of
the qubit systems ($2,3,...,n$) was not involved during the entire
operation, one can choose the qubit systems ($2,3,...,n$) for which the
transition between the two lowest levels $\left| 0\right\rangle $ and $%
\left| 1\right\rangle $ is forbidden or weak to avoid or reduce decoherence
caused by the spontaneous emission from the level $\left| 1\right\rangle .$

The three-level or four-level qubit systems here are widely available in
natural atoms and also in \textit{artificial} atoms such as superconducting
charge-qubit systems [37], phase-qubit systems [38,39], and flux-qubit
systems [37,40]. For a detailed discussion, see Ref. [27].

\begin{center}
\textbf{IV. COMPARING WITH PREVIOUS PROPOSALS}
\end{center}

In this section, we will give a comparison between our proposal and previous
ones. For simplicity, we will compare our proposal with the ones in [17,25].
To the best of our knowledge, the proposals in [17,25] are most closely
related to our work.

\begin{figure}[tbp]
\includegraphics[bb=97 320 464 519, width=8.6 cm, clip]{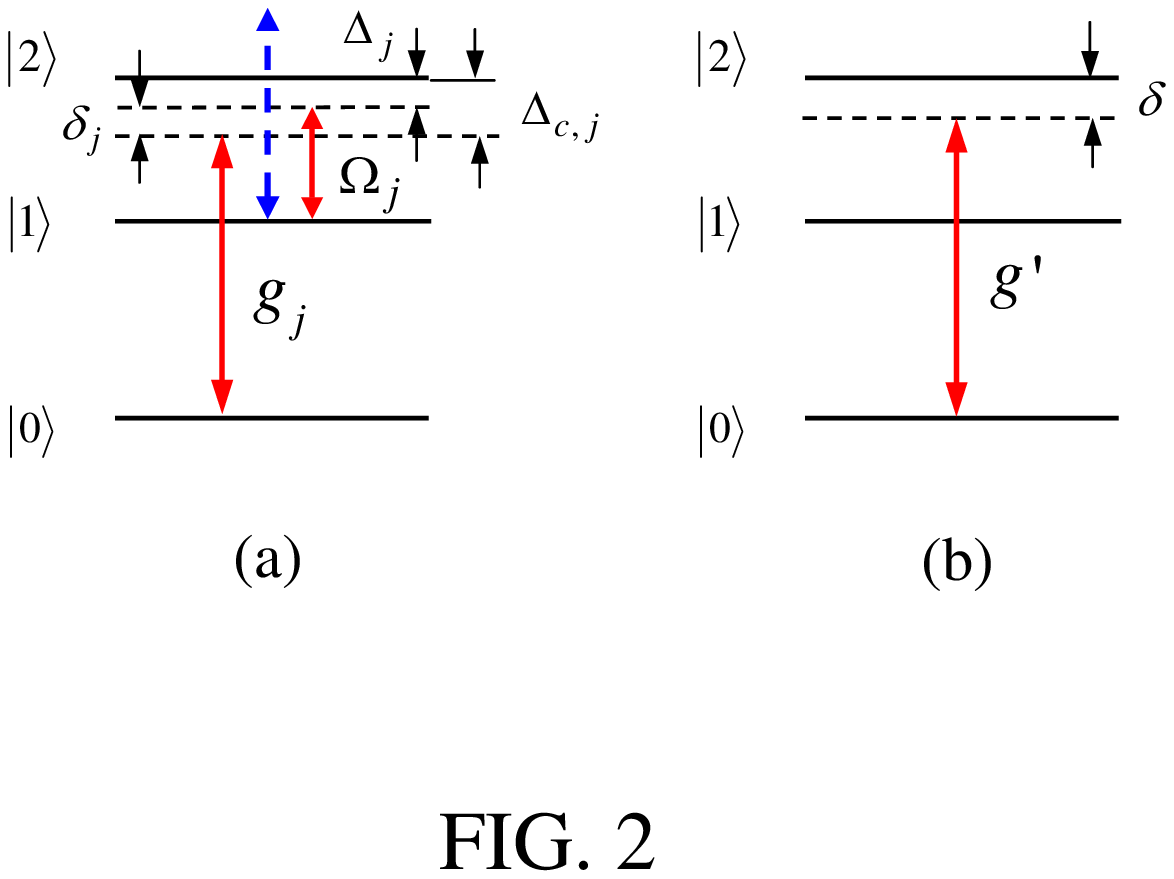} %
\vspace*{-0.08in}
\caption{(Color online) (a) System-cavity-pulse off-resonant Raman coupling
for $n$ qubit systems ($1,2,...,n$) in a cavity. For simplicity, we here
only draw a figure for qubit system $j$ interacting with the cavity mode and
a classical pulse ($j=1,2,...,n$). $\delta _j=\Delta _{c,j}-\Delta _j$ is
the detuning of the cavity mode with the pulse, $\Delta _{c,j}=\omega
_{20}^j-\omega_c $ is the detuning between the cavity mode frequency $%
\omega_c $ and the $\left| 0\right\rangle \leftrightarrow \left|
2\right\rangle $ transition frequency $\omega _{20}^j$ of qubit system $j,$ $%
\Delta _j=\omega _{21}^j-\omega_j $ is the detuning between the pulse
frequency $\omega_j $ and the $\left| 1\right\rangle \leftrightarrow \left|
2\right\rangle $ transition frequency $\omega _{21}^j$ of qubit system $j.$
The blue arrow dashed line represents a second pulse applied to the qubit
system $j$ to cancel the Stark shift of the level $\left| 1\right\rangle$
induced by the first pulse. (b) System-cavity off-resonant interaction for
each of $n$ three-level qubit systems ($1,2,...,n$) in Ref. [17]. $g^{\prime
}$ is the non-resonant coupling strength between the cavity mode and the $%
\left| 0\right\rangle \leftrightarrow \left| 2\right\rangle $ transition. $%
\delta =\omega _{20}-\omega _c$ is the detuning of the cavity mode frequency
with the $\left|0\right\rangle \leftrightarrow \left| 2\right\rangle $
transition frequency.}
\label{fig:2}
\end{figure}

\begin{center}
\textbf{A. Comparison with the proposal in [25]}
\end{center}

An $n$-qubit GHZ state can be prepared without real excitation of the cavity
mode, by using the method introduced in [25]. To see this, consider $n$
three-level qubit systems ($1,2,...,n$) in a cavity. Assume that the cavity
mode is coupled to the $\left| 0\right\rangle \leftrightarrow \left|
2\right\rangle $ transition and the pulses are coupled to the $\left|
1\right\rangle \leftrightarrow \left| 2\right\rangle $ transition, to
establish the system-cavity-pulse off-resonant Raman coupling [Fig.~2(a)].
In this case, we can obtain an effective Hamiltonian, i.e., the Hamiltonian
(10) with a replacement of $\left| 1\right\rangle ,\left| 2\right\rangle ,$
and $\sum_{j=2}^n$ by $\left| 0\right\rangle $, $\left| 1\right\rangle ,$
and $\sum_{j=1}^n$ respectively. When the cavity mode is initially in a
vacuum state $\left| 0\right\rangle _c,$ this effective Hamiltonian reduces
to
\begin{eqnarray}
H_{\mathrm{eff}} &=&-\hbar \sum_{j=1}^n\frac{\Omega _j^2}{\Delta _j}\left|
1\right\rangle _j\left\langle 1\right| +\hbar \sum_{j=1}^n\frac{\chi _j^2}%
\delta \left| 1\right\rangle _j\left\langle 1\right|  \nonumber \\
&&\ \ \ \ \ \ \ \ \ \ \ \ \ \ +\hbar \sum_{j\neq j^{\prime }=1}^n\frac{\chi
_j\chi _{j^{\prime }}}\delta \left( \sigma _{01,j}^{+}\sigma _{01,j^{\prime
}}^{-}+\sigma _{01,j}^{-}\sigma _{01,j^{\prime }}^{+}\right) ,  \nonumber \\
&&
\end{eqnarray}
where $\Delta _j=\omega _{21}^j-\omega _j,$ $\Delta _{c,j}=\omega
_{20}^j-\omega _c,$ and $\delta =\delta_j=\Delta _{c,j}-\Delta _j$ (which
can be reached via adjustment of the pulse frequencies)$.$ The notation of $%
\chi _j$ is the same as that given in Sec. III. To prepare the qubit systems
($1,2,...,n$) in a GHZ state, one will need to apply a second pulse to each
of the qubit systems ($1,2,...,n$) to cancel the Stark shifts, i.e., the
first term of Eq. (25). After that, the Hamiltonian (25) becomes
\begin{eqnarray}
H_{\mathrm{eff}} &=&\hbar \sum_{j=1}^n\frac{\chi _j^2}\delta \left|
1\right\rangle _j\left\langle 1\right| +\hbar \sum_{j\neq j^{\prime }=1}^n%
\frac{\chi _j\chi _{j^{\prime }}}\delta \left( \sigma _{01,j}^{+}\sigma
_{01,j^{\prime }}^{-}+\sigma _{01,j}^{-}\sigma _{01,j^{\prime }}^{+}\right) .
\nonumber \\
&&
\end{eqnarray}
For the case of $\chi _1=\chi _2=...=\chi _n=\chi $ (achievable by changing
the pulse Rabi frequencies), this Hamiltonian is the same as the Hamiltonian
(6) presented in Ref. [25] for $n$ two-level atoms interacting dispersively
with a single-mode cavity. According to the discussion in Ref.~[25], an $n$%
-qubit GHZ state can be prepared based on the Hamiltonian~(26).

From the description here, one can see that to achieve identical Raman
transition strengths $\chi _1,$ $\chi _2,$ $...,$ and $\chi _n$, the pulses
applied to different qubit systems should have different Rabi frequencies if
the qubit-cavity coupling constants are nonidentical. And, to obtain
identical detuning $\delta $ for each qubit system, adjustment of the pulse
frequencies would be needed. Furthermore, to generate GHZ states, a \textit{%
second} off-resonant pulse should be applied to each qubit system to cancel
the Stark shift induced by the first pulse. Namely, for generation of an $n$%
-qubit GHZ state, a total of $2n$ pulses would be required. In contrast, as
shown above, our present proposal, which employs four-level qubit systems,
needs only $\left( n-1\right) $ pulses applied to the qubit systems ($%
2,3,...,n$), i.e., one pulse for each of qubit systems ($2,3,...,n$). Hence,
the use of the pulses is significantly reduced in the present proposal.

\begin{center}
\textbf{B. Comparison with the proposal in [17]}
\end{center}

In Ref.~[17], an auxiliary qubit system $a$ was used to create a single
photon in the cavity mode, which was then employed to prepare the $n$ qubit
systems ($1,2,...,n$) in a GHZ state. As discussed there, measurement on the
states of the auxiliary qubit system $a$ and adjustment of the level
spacings of each qubit system were both needed during the entire operation.
In addition, qubit systems ($1,2,...,n$)\ were required to have identical
three-level structures and the qubit-cavity coupling constants for qubit
systems ($1,2,...,n$) were needed to be the same.

The cavity mode in~[17] was set to be off-resonant resonant with the $\left|
0\right\rangle \leftrightarrow \left| 2\right\rangle $ transition of each of
qubit systems ($1,2,...,n$), with a detuning $\delta =\omega _{20}-\omega _c$
[Fig.~2(b)]. The coupling constant between the cavity mode and the $\left|
0\right\rangle \leftrightarrow \left| 2\right\rangle $ transition for each
of qubit systems ($1,2,...,n$) was denoted as $g^{\prime }.$ As shown in
[17], a multiqubit GHZ state was generated by simultaneously performing a
common phase shift $e^{i\pi }$ on the state $\left| 0\right\rangle $ of each
of qubit systems ($1,2,...,n$) with assistance of the cavity photon, by
having $e^{i\lambda t}=e^{i\pi }$ (i.e., $\lambda t=\pi $). Here, $\lambda
=(g^{\prime })^2/\delta ,$ which was originally defined in~[17]. For a
detailed discussion, see Ref.~[17].

It is noted that the method in~[17] can not work in the case when the
qubit-cavity coupling constants are \textit{nonidentical}. The reason is
that a common phase shift $e^{i\pi }$ can not be simultaneously performed on
the state $\left| 0\right\rangle $ of each of qubit systems ($1,2,...,n$).
For instances, when the qubit-cavity coupling constants $g_j^{\prime }$ and $%
g_k^{\prime }$ for qubits $j$ and $k$ are not the same (resulting in
different $\lambda _j=(g_j^{\prime })^2/\delta $ and $\lambda
_k=(g_k^{\prime })^2/\delta $), one can not have both of $\lambda _jt$ and $%
\lambda _kt$ to be equal to $\pi $ for a given time $t.$ The discussion here
applies to identical atoms in a cavity, for which the detuning $\delta $ of
each atom with the cavity mode is the same but the qubit-cavity couplings
are nonidentical due to nonexact placement of atoms in the cavity.

Let us now see if the method in~[17] can work for a situation that the
detuning of each qubit system with the cavity mode and the qubit-cavity
coupling constants are both nonidentical. This situation applies to
superconducting qubit systems in a cavity. As discussed above, to prepare a
multiqubit GHZ state, the condition $\lambda _j=\lambda _k$ needs to be
satisfied for arbitrary two superconducting qubit systems $j$ and $k$. Here,
$\lambda _j=(g_j^{\prime })^2/\delta _j$ and $\lambda _k=(g_k^{\prime
})^2/\delta _k.$ Note that once the superconducting qubit systems are
designed and the cavity mode frequency is chosen, the detunings $\delta _j$
and $\delta _k$ are fixed. Thus, in order to obtain $\lambda _j=\lambda _k$,
i.e., $(g_j^{\prime })^2/\delta _j=(g_k^{\prime })^2/\delta _k,$ it would be
required to exactly place the qubit systems in the cavity to have \textit{%
desired} qubit-cavity coupling constants $g_j^{\prime }$ and $g_k^{\prime },$
which is not easy to achieve in experiments due to the fabrication error.
This problem becomes more apparent especially when the number of qubit
systems in a cavity is large. In contrast, as shown above, this difficulty
is avoided in our present proposal, because the effective coupling strength $%
\lambda _j$ in Eq.~(13) can be adjusted by changing the intensity of the
pulses, instead of exactly placing qubit systems in a cavity to have desired
qubit-cavity coupling constants.

From the discussion here, it can be concluded that our present proposal is
quite different from the previous ones [17,25]. Due to the use of a
four-level structure, the present proposal can be used to prepare a
multiqubit GHZ state for nonidentical qubit-cavity coupling constants
(compared with the proposal in [17]), and requires less application of
pulses (compared with the proposal in [25]).

\begin{center}
\textbf{V. DISCUSSION}
\end{center}

In this section, we will give a brief discussion on the experimental issues.
For the method to work:

(i) The occupation probability $p_j$ of the level $\left| 3\right\rangle $
for qubit system $j$ ($j=2,3,...,n$) during step (ii) is given by [28]
\begin{equation}
p_j\simeq \frac 2{4+\left( \Delta _j/\Omega _j\right) ^2}+\frac 2{4+\left(
\Delta _{c,j}/g_j\right) ^2},
\end{equation}
which needs to be negligibly small in order to reduce the operation error.

(ii) As discussed above, the conditions (21), (22), and (23) need to be
satisfied;

(iii) The total operation time $\tau $ given in Eq.~(24) should be much
shorter than the energy relaxation time $\gamma _{1r}^{-1}$ and dephasing
time $\gamma _{1p}^{-1}$ of the level $\left| 1\right\rangle $ and the
lifetime of the cavity mode $\kappa ^{-1}=Q/2\pi \nu _c,$ where $Q$ is the
(loaded) quality factor of the cavity.

The above requirements can in principle be realized, since one can: (i)
reduce $p_j$ by increasing the ratio of $\Delta _j/\Omega _j$ and $\Delta
_{c,j}/g_j;$ (ii) shorten $t_{1a},t_{1b},t_{1c}$ by increasing the pulse
Rabi frequency $\Omega _r,$ $\widetilde{\Omega }_r$, and the coupling
constant $g$; (iii) design qubit systems (e.g., superconducting devices) to
have sufficiently long energy relaxation times $\gamma _{1r}^{-1}$ and $%
\gamma _{2r}^{-2}$ and dephasing times $\gamma _{1p}^{-1}$ and $\gamma
_{2p}^{-1}$ or choose qubit systems (e.g., atoms) with long decoherence
times of the levels $\left| 1\right\rangle $ and $\left| 2\right\rangle ;$
(iii) increase $\kappa ^{-1}$ by employing a high-$Q$ cavity so that the
cavity dissipation is negligible during the operation.

\begin{figure}[tbp]
\includegraphics[bb=66 166 560 627, width=10.6 cm, clip]{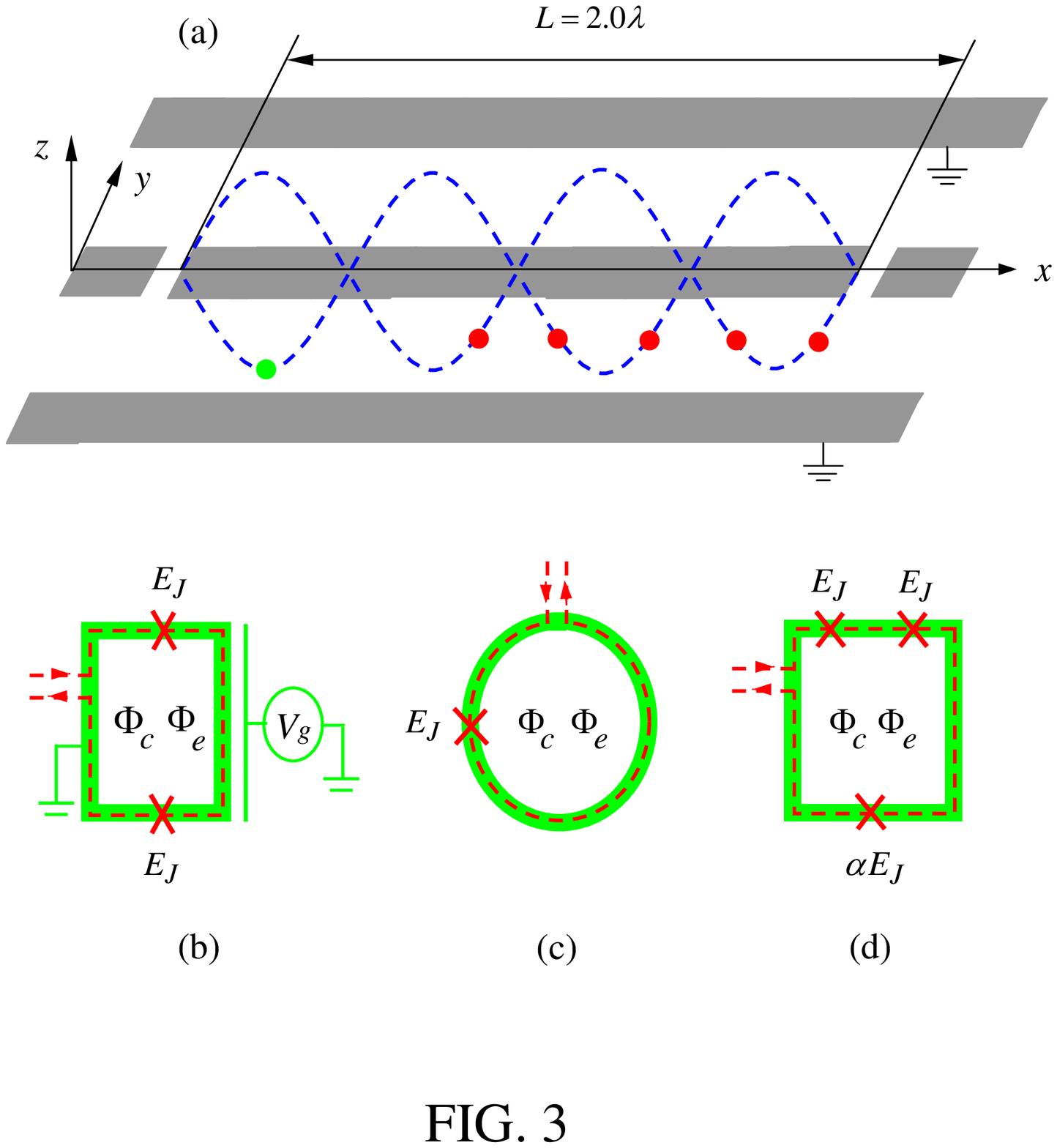} %
\vspace*{-0.08in}
\caption{(Color online) (a) Sketch of the setup for six superconducting
qubit systems and a (grey) standing-wave quasi-one-dimensional coplanar
waveguide resonator. The two blue curved lines represent the standing wave
magnetic field, which is in the $z$-direction. The green dot represents the
qubit system 1, which is placed at an antinode of the standing wave magnetic
field to achieve the maximal qubit-cavity coupling constant $g$. The red
dots represent the qubit systems $2,3,...,$ and $n$, which are placed at
locations where the magnetic fields are the same. Each qubit system could be
a superconducting charge-qubit system shown in (b), flux-biased phase-qubit
system in (c), and flux-qubit system in (d). The superconducting loop of
each qubit system, which is a large square for (b) and (d) while a large
circle for (c), is located in the plane of the resonator between the two
lateral ground planes (i.e., the $x$-$y$ plane). Each qubit system is
coupled to the cavity mode through a magnetic flux $\Phi_c$ threading the
superconducting loop, which is created by the magnetic field of the cavity
mode. A classical magnetic pulse is applied to each qubit system through an
AC flux $\Phi_e$ threading the qubit superconducting loop, which is created
by an AC current loop (i.e., the red dashed-line loop) placed on the qubit
loop. The frequency and intensity of the pulse can be adjusted by changing
the frequency and intensity of the AC loop current. $E_{J}$ is the Josephson
junction energy ($0.6<\alpha<0.8 $) and $V_g$ is the gate voltage. $\lambda$
is the wavelength of the resonator mode, and $L$ is the length of the
resonator.}
\label{fig:3}
\end{figure}

For the sake of definitiveness, let us consider the experimental possibility
of generation of a six-qubit GHZ state with six superconducting qubits
coupled to a resonator (Fig.~3). With the choice of $\Omega _r,$ $\widetilde{%
\Omega }_r\sim 10g$, we have $t_{1a},t_{1c}\sim 0.05\pi /g$ and $t_{1b}\sim
0.5\pi /g.$ By setting $\Delta _j\sim 10\Omega _j,$ $\Delta _{c,j}\sim
10g_j, $ and $\Omega _j\sim 0.9g_j,$ we have $\delta =\delta _j\sim
10g_j^2/\Delta _{c,j}\sim 10\Omega _j^2/\Delta _j\sim 10\chi _j,$ leading to
$\lambda _j\sim 0.11g_j.$ For $g_j\sim 0.2g,$ we can set $\lambda _2=\lambda
_3=...=\lambda _6=$ $\lambda \sim 0.022g$, which can be achieved by
adjusting the Rabi frequencies $\Omega _2,$ $\Omega _3,...,\Omega _6$ of the
pulses applied to the qubit systems ($2,3,...,6$) as discussed above. Thus,
we have $\tau \sim 46.6\pi /g.$ As a rough estimate, assume $g/{2\pi }\sim
220$ MHz, which could be reached for a superconducting qubit system coupled
to a one-dimensional standing-wave CPW (coplanar waveguide) transmission
line resonator [42]. For the $g$ chosen here, we have: (i) $%
t_{1a},t_{1c}\sim 0.1$ ns and $t_{1b}\sim 1.1$ ns, which shows that the
condition (22) is satisfied; (ii) $t_{1a}+t_{1b}\sim 1.2$ ns, showing that
the condition (23) is met for $\min \{\gamma _{2r}^{-1},\gamma
_{2p}^{-1}\}\sim 0.5$ $\mu $s [38,43]$;$ and (iii) $\tau \sim 0.1$ $\mu $s,
much shorter than $\min \{\gamma _{1r}^{-1},\gamma _{1p}^{-1}\}\sim 1$ $\mu $%
s [38,43]. In addition, consider a resonator with frequency $\nu _c\sim 3$
GHz (e.g., Ref. [44]) and $Q\sim 5\times 10^4$, we have $\kappa ^{-1}\sim
2.5 $ $\mu $s, which is much longer than the total operation time $\tau $.
Note that superconducting coplanar waveguide resonators with a quality
factor $Q>10^6$ have been experimentally demonstrated [45].

For the choice of $\Delta _j\sim 10\Omega _j$ and $\Delta _{c,j}\sim 10g_j$
here, we have $p_j\sim 0.04$, which can be further reduced by increasing the
ratio of $\Delta _j/\Omega _j,$ and $\Delta _{c,j}/g_j.$ In addition, for
the parameters chosen here, when the unwanted phase shifts mentioned above
are considered, a simple calculation shows that the fidelity $\emph{F}$ is $%
\sim 0.992.$ We should mention that further investigation is needed for each
particular experimental set-up. However, this requires a rather lengthy and
complex analysis, which is beyond the scope of this theoretical work.

As shown in Sec.~III, this proposal requires simultaneous application of
pulses with different frequencies and intensities during the operation of
step (ii). For the set up in Fig. 3(a), simultaneous application of several
pulses of different frequencies and intensities to qubit systems $2,3,4,$
and $5$ is feasible in experiments [46]. According to Ref. [46], for a qubit
system containing a superconducting loop, one can place an AC current loop
on the qubit superconducting loop to create an AC flux $\Phi _e$ (i.e., a
classical magnetic pulse) threading the qubit loop [Fig.~3(b,c,d)]; and
the frequencies and intensities of pulses applied to qubit systems $%
2,3,4,$ and $5$ can be readily and simultaneously changed by varying the
frequencies and intensities of the AC loop currents of the qubit systems $%
2,3,4,$ and $5$ at the same time.

\begin{center}
\textbf{VI. CONCLUSION}
\end{center}

In summary, we have proposed a way for creating $n$-qubit GHZ entangled
states with a three-level qubit system and $\left( n-1\right) $ four-level
qubit systems in a cavity or coupled to a resonator. The main advantage of
this proposal is that no identical qubit-cavity coupling constants are
required. As a result, qubit systems (e.g., solid-state qubits), which often
have parameter nonuniformity, can be used; and no exact placement of qubits
in a cavity is needed. Note that for solid-state qubits (e.g.,
superconducting qubits), it is difficult to design qubits with identical
device parameters; and it is experimentally challenging to place many qubits
at different locations where the magnetic fields or electric fields are
exactly the same. These hardships are avoided in this proposal. Another
interesting property of this proposal is that during the entire operation,
there is no need of adjusting the level spacings of the qubit systems or the
cavity mode frequency. In experiments, adjustment of the qubit level
spacings or the cavity mode frequency during the operation is not desired;
and extra errors may be induced by adjustment of the qubit level spacings or
the cavity mode frequency. Furthermore, as shown above, this proposal has
these additional advantages: (i) The $n$-qubit GHZ state is prepared
deterministically and the operation time is independent of the number of
qubits in the cavity; (ii) Neither auxiliary qubit systems nor measurement
on the qubit states is needed; and (iii) No photon detection is needed. This
proposal is quite general, which can be applied to various types of
superconducting qubits and atoms trapped in a cavity.

\begin{center}
\textbf{ACKNOWLEDGMENTS}
\end{center}

C.P.Y. is grateful to Shi-Biao Zheng for very useful comments. This work is
supported in part by the National Natural Science Foundation of China under
Grant No. 11074062, the Zhejiang Natural Science Foundation under Grant No.
Y6100098, funds from Hangzhou Normal University, and the Open Fund from the
SKLPS of ECNU.

\end{document}